\documentclass[runningheads]{llncs}
\usepackage[T1]{fontenc}
\usepackage{graphicx,verbatim}
\usepackage{amsmath}
\usepackage{amssymb}
\usepackage{booktabs}
\usepackage{multirow}

\usepackage{hyperref}
\usepackage{color}

\usepackage{orcidlink}

\begin{document}
\title{Multi-Stage NeRF for Efficient 3D Coronary Artery Reconstruction from Two Narrow-Angle Angiographic Projections}
\titlerunning{Multi-Stage NeRF for 3D CA Reconstruction from Two Angiograms}
%

\author{Deyu Meng \orcidlink{0009-0002-0659-827X} \and
Mojtaba Lashgari \orcidlink{0000-0002-5307-3495} \and
Yiying Wang \orcidlink{0009-0007-6572-5474} \and
Abhirup Banerjee \orcidlink{0000-0001-8198-5128}}  
\authorrunning{D. Meng et al.}
\institute{
Institute of Biomedical Engineering, Department of Engineering Science,\\
University of Oxford, Oxford, United Kingdom\\
    \email{deyu.meng@eng.ox.ac.uk}}
  
\maketitle              
\begin{abstract}

X-ray coronary angiography is the clinical gold standard for coronary artery disease during real-time cardiac interventions, but provides only 2D projections of inherently 3D vessels. Existing learning-based 2D-to-3D reconstruction methods typically require wide angular coverage or multiple views, assumptions that are rarely satisfied in routine practice where only two projections with narrow angular separation are available.
To address these challenges, we propose NeCA++, a multi-stage self-supervised neural radiance field (NeRF) framework tailored to clinically realistic acquisition constraints. 
The framework decomposes reconstruction into two stages that progressively refine spatial support and representation capacity. In the first stage, a coarse 3D representation of the vasculature is reconstructed, restricting the subsequent optimisation to regions with a higher likelihood of vessel presence, termed an active region. Afterward reconstruction is restricted to this region while higher-resolution representations are progressively activated to recover fine vascular details.
This multi-stage strategy focuses learning on anatomically plausible regions, mitigates gradient dilution under extreme sparsity, and stabilises global topology before recovering fine vascular branches.
Furthermore, two vessel-specific regularisations are introduced: a ray-aligned constraint to reduce projection-induced ambiguity, and a bimodal density penalty to enable early vessel-background separation.
Extensive experiments across three datasets (ImageCAS, ASOCA, and Synthetic RCA) and four angular configurations demonstrate consistent superiority over state-of-the-art baselines, particularly under clinically realistic narrow-angle settings, while achieving reconstruction within 58 seconds per case. The code is available at \href{\detokenize{https://github.com/MultiMeDIA-Oxford/NeCAPP}}{https://github.com/MultiMeDIA-Oxford/NeCAPP}.


\end{abstract}
%
%
\section{Introduction}

Coronary artery disease (CAD) remains a leading cause of morbidity and mortality worldwide \cite{townsend2022epidemiology,Virani2021HeartDisease}, highlighting the need for accurate diagnosis and effective intervention. 
X-ray coronary angiography is the gold standard for procedural guidance during coronary interventions and is one of the principal imaging modalities for the assessment of CAD, owing to its high spatial and temporal resolution and real-time visualisation of the coronary lumen \cite{lashgari2024patient}.
However, it provides only two-dimensional (2D) images of three-dimensional (3D) vascular structures, resulting in vessel overlap and foreshortening. Clinicians subjectively infer the 3D structure of arteries from limited views, which can affect stenosis assessment and lesion localisation \cite{green2005angiographic}. Automated 3D coronary artery (CA) reconstruction improves spatial understanding, enabling effective diagnosis and intervention.

Early 3D CA reconstruction methods are predominantly based on computational geometry, recovering vessel centerlines \cite{blondel2006reconstruction,galassi20183d,jandt2009automatic,unberath2017symmetry} or luminal surfaces \cite{banerjee2019point,chen2009three,ccimen2016reconstruction,schoonenberg2009three,wiesent2000enhanced} from multiple projections through explicit inter-view correspondences. Although these approaches can achieve high reconstruction accuracy under controlled settings, 
most of these methods require manual annotation
and exhibit limited scalability and substantial computational time requirements. Learning-based methods, in contrast, mitigate these dependencies by automatically estimating model parameters from data. 
For example, DeepCA \cite{wang2025deepca} employs a generative adversarial network (GAN) \cite{goodfellow2020generative} for 3D CA reconstruction.
However, such supervised approaches rely on a substantial amount of paired 2D angiographic projections and corresponding 3D ground truth (GT) data.
In another approach, neural radiance field (NeRF) \cite{mildenhall2021nerf} introduces a self-supervised implicit representation that enables 3D reconstruction without case-specific annotations. 
NeRF-CA \cite{maas2025nerf} and its improved version NeRT-CA \cite{maas2025nert} extend NeRF to dynamic reconstruction with emphasis on temporal consistency, reconstructing CA from four and three projections, respectively.
In contrast, NeCA \cite{wang2024neca}, also built on NeRF, addresses static reconstruction from two projections, but assumes a near-orthogonal angle separation between them. 
Nevertheless, the aforementioned methods differ from routine clinical practice, where only two projections with narrow angular separation are typically available.

To address these challenges under clinically constrained narrow-angle acquisition, we propose NeCA++, a computationally efficient self-supervised multi-stage neural implicit framework for 3D CA reconstruction from two narrow-angle projections. Rather than directly optimising a high-capacity model over a highly sparse volume in 3D space, NeCA++ decomposes optimisation into two stages that progressively refine spatial support and representation capacity. A coarse 3D vascular structure is reconstructed in the first stage to limit optimisation to a 3D subspace, termed an active region, using low-resolution hash encoding. The subsequent optimisation is restricted to this region, while higher-resolution hash levels are progressively activated to refine structural details. This staged design concentrates learning on anatomically plausible subspaces, mitigates gradient dilution under extreme sparsity, and stabilises topology before recovering fine vascular branches while speeding up the reconstruction. In addition, novel regularisations further constrain the solution under limited angular diversity. Extensive experiments are conducted under clinically realistic narrow-angle acquisition protocols, a setting that remains under-explored in prior reconstruction studies. Results demonstrate robust performance over the ImageCAS, ASOCA, and Synthetic RCA datasets with significant improvements over state-of-the-art methods and an average runtime of 58 seconds per case.

Our main contributions are:
\textbf{1)} A multi-stage neural implicit framework that localises an initial region of vascular presence and progressively reconstructs 3D coronary arterial tree based on a coarse-to-fine approach, for two projections with narrow-angle separation;
\textbf{2)} Two vessel-specific regularisation strategies, including a ray-aligned geometry constraint and a bimodal density penalty, explicitly designed to mitigate shape inference ambiguity and facilitate early vessel-background separation; and
\textbf{3)} Extensive validation under clinical narrow-angle acquisition protocols, demonstrating strong robustness and efficiency.

\section{Method}
\label{sec:method}

This section presents the proposed NeCA++ for 3D coronary tree reconstruction from two projections with narrow-angle separation. Section~\ref{sec:network} describes the overall structure of NeCA++.
Section~\ref{sec:training} describes a two-stage optimisation strategy where the first stage estimates a coarse 3D reconstruction of the vascular tree to isolate a highly likely 3D subspace where the vascular tree is present, termed an active region.
Then, the second stage progressively refines predictions within the active region. 
Finally, Sec. \ref{sec:loss_new} explains the vessel-specific regularisation terms in NeCA++ in order to reduce projection ambiguity and vessel-background separation under narrow angular separation.

\begin{figure}[t]
    \centering
    \includegraphics[width=12cm,height=5.2cm]{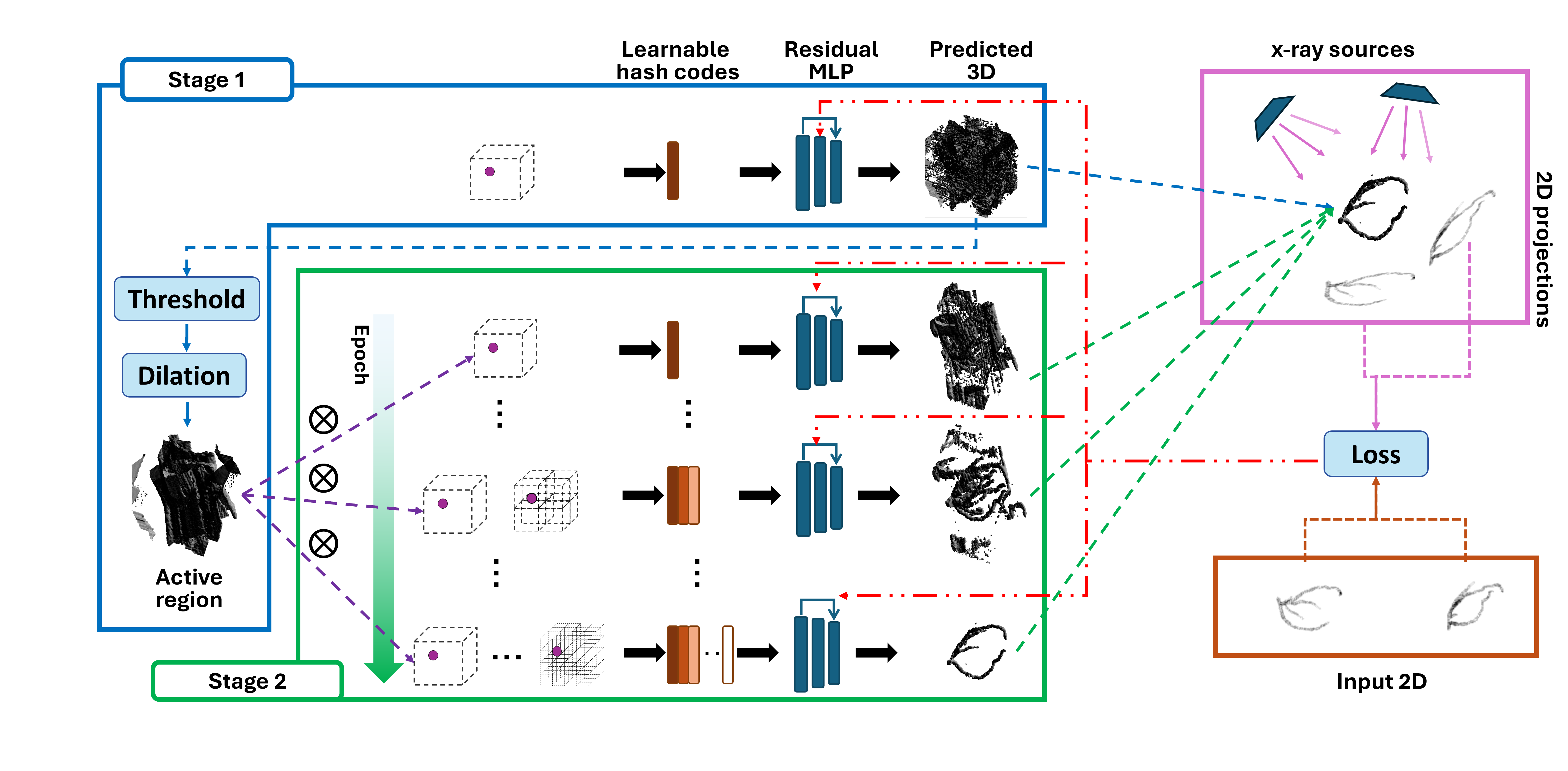}
    \caption{Overview of the proposed NeCA++, the multi-stage NeRF framework. 
NeCA++ reconstructs 3D coronary geometry from two narrow-angle X-ray angiographic projections using a self-supervised neural implicit field. In Stage 1, NeCA++ localises an active region for optimisation from a coarse 3D reconstruction of the vasculature. In Stage 2, the coarse 3D reconstruction is optimised progressively by activating higher-resolution hash encoders to refine topology and recover fine vascular branches, while the optimisation is restricted to the active region.
 }
    \label{fig:net}
\end{figure}

\subsection{Implicit Field Representation and Reconstruction Network}
\label{sec:network}
The proposed NeCA++ follows the neural field formulation of NeRF and reconstructs the unknown 3D structure of a vascular tree from two X-ray angiography projections $Y_1, Y_2$ acquired with narrow angular separation \cite{green2005angiographic}.
Let $\Omega \subset \mathbb{R}^3$ denote a bounded 3D space discretised into $N$ voxel centers
$\{\mathbf{p}_i\}_{i=1}^{N}$.
The corresponding 2D projections of a real vascular tree, acquired using X-ray angiography, satisfy the following image formation:
$
\mathbf{c}_{1,i} = \mathbf{P}_1 \mathbf{p}_i, 
\mathbf{c}_{2,i} = \mathbf{P}_2 \mathbf{p}_i,
$
where $\mathbf{p}_i$,
$\mathbf{c}_{1,i}$, and $\mathbf{c}_{2,i}\in\mathbb{R}^{3}$ are image coordinates, and
$\mathbf{P}_1,\mathbf{P}_2$ are $3\times3$ projection matrices associated with image planes $Y_1$ and $Y_2$.
The objective is to estimate a vascular occupancy value at each spatial location $\textbf{p}_i \in {\Omega}$.

A multi-resolution hash encoding inspired by InstantNGP~\cite{muller2022instant} is applied on 
a general 3D space of  $\Omega$ with the same $\textbf{p}_i \in {\Omega}$, to enhance representation capacity and accelerate convergence.
Features are extracted at $L$ resolution levels.
For each level $\ell$, a hash table stores corresponding feature vectors as learnable parameters $\Phi$.
The input coordinate is mapped to the $\ell$-th grid to locate the surrounding voxel vertices.
Their coordinates are converted into hash indices to access the corresponding feature vectors.
The retrieved vertex features are trilinearly interpolated at each level, and the resulting features from all $L$ levels are concatenated to form the multi-resolution encoding $\gamma(x)$.
Then, this encoded feature $\gamma(x)$ is mapped to a vascular occupancy value by a coordinate-based multilayer perceptron (MLP) $f_\theta$. 
For each voxel centre $\textbf{p}_i$, this gives $\sigma_\theta(\textbf{p}_i)=\mathrm{sigmoid}(f_\theta(\gamma(\textbf{p}_i)))$, and the collection of these values over all voxels forms the reconstructed $V$.

To optimise reconstruction of an accurate 3D vascular structure in NeCA++, we enforce consistency with 2D input projections, $Y_v$, through $L_{\text{proj}}$ along with two other regularisations explained in Sec.~\ref{sec:loss_new}.
With $L_{\text{proj}}$, first, 2D projection of the reconstructed volume, $\hat{V}_{\theta}$, is rendered using a differentiable cone-beam projector implemented with TIGRE~\cite{biguri2025tigre}, as highlighted in Fig.~\ref{fig:net} (pink box).
Then, the rendered projection $\hat{Y}_v$ is compared with the observed image $Y_v$ (brown box in Fig.~\ref{fig:net}), to drive optimisation. The two-view projection loss is defined as
\begin{equation}
L_{\text{proj}}
=
\sum_{v\in\{1,2\}}
\|
\hat{Y}_v - Y_v
\|_2^2.
\end{equation}

Unlike conventional learning-based approaches that train a network across multiple training cases, the NeCA++ optimises network parameters specifically for each pair of 2D projections at the input.

\subsection{Multi-stage Optimisation of Hash Encoding}
\label{sec:training}

Vascular structures occupy only a small fraction of $\Omega$. As a result, generating 2D projections of the 3D reconstructed vessels to calculate $L_{\text{proj}}$ from the entire $\Omega$ is inefficient. 
Instead, concentrating the optimisation on a subspace with more likelihood of vascular presence, termed the active region $\Omega^{*}$, avoids unnecessary computation associated with background voxels, leading to fast convergence. 
To exclude background voxels from computation, we adopt a two-stage optimisation strategy (Fig.~\ref{fig:net}). The first stage localises the active region, and the second refines predictions within this region while progressively increasing representational capacity to reconstruct finer details of vascular structure.

In the \textit{first stage}, the network is optimised over the entire $\Omega$ and reconstructs a coarse structure of the vascular tree, $\hat{V}_\theta$ (blue box in Fig.~\ref{fig:net}). 
 $\hat{V}_\theta$ is then binarised to define $\Omega^{*}$.
To reduce the risk of missing thin branches, a $3\times3\times3$ morphological dilation is applied to $\hat{V}_\theta$ to obtain $\Omega^{*}$ subspace.
The first four low hash encoding levels are used in this stage, and the learned weight and low-resolution hash parameters are retained to initialise the second stage. 

In the \textit{second stage}, optimisation proceeds in a progressive manner. Here, the number of active hash encoding levels increases monotonically from the lowest level to higher-resolution hash levels. Through this multi-scale procedure, at the low hash levels, the global topology of the vascular structure is estimated. Then, by increasing the hash levels, finer levels are gradually introduced to sharpen boundaries and recover thin vascular branches.

\subsection{Vessel-specific Structural Regularisation}


\label{sec:loss_new}

The projection loss $L_{\text{proj}}$ introduced in Sec.~\ref{sec:network} cannot enforce accurate 3D shape reconstruction and separation between background and vascular structure due to overlaying the information of voxels along the X-ray beam at a pixel. 
For this purpose, two regularisation terms of $L_{\text{ray}}$ and $L_{\text{bin}}$ are introduced to enforce accurate 3D shape reconstruction and separation between background and vascular structure, respectively. 
To calculate $L_{\text{ray}}$, first we derive the X-ray beam direction from the known source to detector as
$\hat{r}$. The gradients of the voxels along the beams are calculated as follows:
\begin{equation}
L_{\text{ray}} =
\frac{1}{N}
\sum_{v}\sum_{j}\sum_{k}
\frac{\partial \sigma_\theta(p_k)}
{\partial \hat r_{vj}},
\end{equation}
where $\sum_{v}$ is the summation over different views $v$, $\sum_{j}$ is the summation over all beams in each view, $\sum_{k}$ is the summation over all voxel centres $p_k$ along each beam, and $N$ is the total number of voxels.

In addition, to encourage separation between background and vessel in $\hat{V}_\theta$, we increase the influence of $\sigma_\theta$ with values close to 0 and 1, while suppressing influence of ambiguous $\sigma_\theta$, i.e. $\sigma_\theta \rightarrow 0.5$, using a bimodal density penalty:
\begin{equation}
L_{\text{bin}} =
\frac{1}{N}
\sum_{i}
\sigma_\theta(\textbf{p}_i)
\Bigl(1-\sigma_\theta(\textbf{p}_i)\Bigr),
\end{equation}
where $\textbf{p}_i \in \Omega^{*}$.
This encourages earlier vessel-background separation during optimisation, which typically reduces the required training iterations. When used together with the ray-aligned constraint, the binarisation effect can improve reconstruction quality by making gradient anomalies more explicit.

The overall loss function is calculated as:
\begin{equation}
L =
L_{\text{proj}}
+
\lambda_{\text{ray}} L_{\text{ray}}
+
\lambda_{\text{bin}} L_{\text{bin}},
\end{equation}
where $\lambda_{\text{ray}}=7.5$ and $\lambda_{\text{bin}}=2$ are weighting factors experimentally selected for the regularisation terms.

\section{Experiments and Results}

\subsection{Experimental Settings}

\paragraph{\textbf{Datasets and Angle Configuration.}} 
Experiments are conducted on three clinical and synthetic 3D CA datasets, including the ImageCAS dataset ($N=40$) \cite{zeng2023imagecas} and ASOCA dataset ($N=40$) \cite{gharleghi2022automated}, each comprising segmented left coronary arteries (LCA) and right coronary arteries (RCA). 
In addition, 100 synthetic RCA cases are generated following~\cite{iyer2023multi}. 
Consistent with prior works~\cite{maas2025nert,maas2025nerf,wang2025deepca,wang2024neca}, 2D X-ray projections are generated from 3D volumes using TIGRE~\cite{biguri2025tigre}.

X-ray acquisition follows a standard C-arm configuration parameterised by (primary angle, secondary angle). 
Guided by clinical angiography protocols~\cite{green2016optimal}, we evaluate four angular separations for each artery type:
RCA: orthogonal (0,0)/(0,90); large (40,-20)/(0,30); medium (40,0)/(0,30); small (40,0)/(20,20); 
LCA: orthogonal (0,0)/(0,90); large (-30,0)/(40,30); medium (-30,0)/(0,40); small (-30,0)/(-40,30).

\paragraph{\textbf{Baselines, Metrics, and Implementation.}} 
We compare NeCA++ with three state-of-the-art 3D CA reconstruction models: NeRT-CA \cite{maas2025nert}, DeepCA \cite{wang2025deepca}, and NeCA \cite{wang2024neca}. As DeepCA is supervised, 75\% of the data are used for fine-tuning and the remaining 25\% for testing. Since DeepCA was originally designed for RCA reconstruction, its evaluation is restricted to RCA cases for fair comparison. Reconstructed 3D volumes are thresholded into binary masks for evaluation. Dice, intersection over union (IoU), and centerline Dice (clDice) \cite{shit2021cldice} are computed against GT 3D models. Due to GPU memory constraints, all experiments use a resolution of $80 \times 80 \times 80$ and are conducted on an Intel Ultra~9 275HX CPU with a single 16GB NVIDIA RTX 5080 GPU. 
All hyperparameters were selected based on the performance of the network over the validation set through sensitivity analysis. The hyperparameter settings in which the performance remained stable across reasonable parameter variations (<2.3\% Dice difference) were used for all datasets and angular configurations.

\subsection{Results}

\begin{figure}[t]
    \centering
    \includegraphics[width=0.9\textwidth]{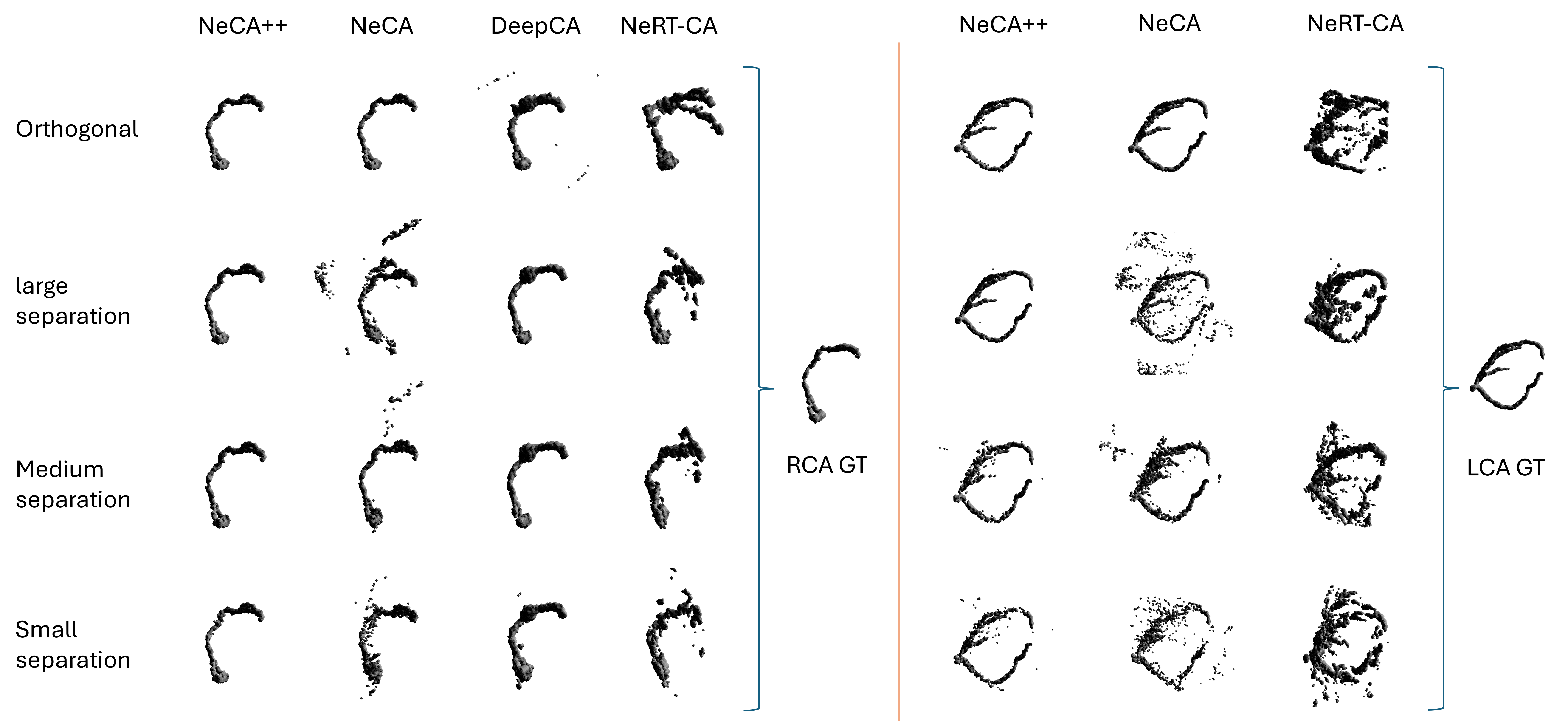}
    \caption{3D coronary artery reconstruction results on ImageCAS dataset, demonstrating significant outperformance of NeCA++ over state-of-the-art approaches. Designed for RCA reconstruction, DeepCA is evaluated only on RCA data.
 }
    \label{fig:result}
\end{figure}

\begin{table*}[t]
\centering
\caption{3D reconstruction performance under different X-ray angular settings. Dice, IoU, and clDice are computed for each dataset.}
\label{tab:compact_recon}
\fontsize{8}{9}\selectfont
\setlength{\tabcolsep}{1pt} 
\renewcommand{\arraystretch}{0.95} 

\begin{tabular}{@{} l  *{4}{ccc} @{}}
\toprule
\multirow{2}{*}{Method} 
  & \multicolumn{3}{c}{Orthogonal} 
  & \multicolumn{3}{c}{Large} 
  & \multicolumn{3}{c}{Medium} 
  & \multicolumn{3}{c}{Small} \\
\cmidrule(lr){2-4}\cmidrule(lr){5-7}\cmidrule(lr){8-10}\cmidrule(lr){11-13}
 & Dice$\uparrow$ & IoU$\uparrow$ & clDice$\uparrow$ & Dice$\uparrow$ & IoU$\uparrow$ & clDice$\uparrow$ & Dice$\uparrow$ & IoU$\uparrow$ & clDice$\uparrow$ & Dice$\uparrow$ & IoU$\uparrow$ & clDice$\uparrow$ \\
\midrule
\multicolumn{13}{c}{ImageCAS Dataset} \\
\midrule
NeRT-CA  & 0.582 & 0.410 & 0.486 & 0.581 & 0.410 & 0.657 & 0.537 & 0.367 & 0.526 & 0.451 & 0.291 & 0.424 \\
NeCA     & \textbf{0.966} & \textbf{0.937} & \textbf{0.954} & 0.569 & 0.391 & 0.437 & 0.688 & 0.529 & 0.606 & 0.568 & 0.403 & 0.500 \\
DeepCA   & 0.582 & 0.406 & 0.746 & 0.593 & 0.422 & 0.822 & 0.593 & 0.421 & 0.800 & 0.628 & 0.457 & \textbf{0.904} \\
NeCA++     & 0.909 & 0.846 & 0.871 & \textbf{0.917} & \textbf{0.854} & \textbf{0.888} & \textbf{0.842} & \textbf{0.741} & \textbf{0.817} & \textbf{0.777} & \textbf{0.652} & 0.752 \\
\midrule

\multicolumn{13}{c}{ASOCA Dataset} \\
\midrule
NeRT-CA  & 0.631 & 0.461 & 0.510 & 0.576 & 0.405 & 0.664 & 0.472 & 0.309 & 0.525 & 0.538 & 0.368 & 0.572 \\
NeCA     & \textbf{0.995} & \textbf{0.990} & \textbf{0.996} & 0.618 & 0.451 & 0.517 & 0.773 & 0.640 & 0.737 & 0.698 & 0.542 & 0.652 \\
DeepCA   & 0.736 & 0.582 & 0.946 & 0.734 & 0.580 & 0.844 & 0.762 & 0.616 & \textbf{0.956} & 0.756 & 0.607 & \textbf{0.943} \\
NeCA++ & 0.992 & 0.985 & 0.989 & \textbf{0.986} & \textbf{0.973} & \textbf{0.981} & \textbf{0.941} & \textbf{0.893} & 0.946 & \textbf{0.932} & \textbf{0.875} & 0.925 \\
\midrule
\multicolumn{13}{c}{Synthetic RCA Dataset} \\
\midrule
NeRT-CA  & 0.623 & 0.452 & 0.638 & 0.547 & 0.376 & 0.565 & 0.526 & 0.312 & 0.505 & 0.610 & 0.439 & 0.656 \\
NeCA     & 0.990 & 0.981 & \textbf{0.995} & 0.582 & 0.414 & 0.444 & 0.735 & 0.585 & 0.645 & 0.642 & 0.475 & 0.557 \\
DeepCA   & 0.560 & 0.389 & 0.798 & 0.660 & 0.493 & \textbf{0.923} & 0.755 & 0.607 & \textbf{0.960} & 0.718 & 0.560 & \textbf{0.969} \\
NeCA++ & \textbf{0.993} & \textbf{0.986} & 0.993 & \textbf{0.930} & \textbf{0.872} & 0.902 & \textbf{0.914} & \textbf{0.845} & 0.910 & \textbf{0.880} & \textbf{0.790} & 0.894 \\
\bottomrule
\end{tabular}
\end{table*}

Reconstruction results are evaluated both qualitatively and quantitatively. 
Figure~\ref{fig:result} shows representative 3D reconstructions. 
NeCA achieves near-perfect performance under orthogonal views but degrades rapidly as angular separation decreases, producing substantial outliers. 
DeepCA is less sensitive to angular settings but yields incorrect surface geometry, particularly around thin branches. 
NeRT-CA exhibits discontinuities and spurious vessels under two inputs. 
In contrast, our proposed NeCA++ preserves vascular geometry across both large and small angular separations, maintaining stable performance with limited outliers.

Table~\ref{tab:compact_recon} reports 3D reconstruction performance under four pairs of angular separation (Orthogonal, Large, Medium, and Small) across three datasets. The data for DeepCA comes from the RCA results, while the other models were tested on both RCA and LCA.
Under the Orthogonal setting, where angular separation is sufficient, NeCA achieves robust reconstruction on all datasets, whereas our proposed NeCA++ achieves comparable performance (Dice 0.909 on ImageCAS, 0.992 on ASOCA, and 0.993 on Synthetic RCA).
As angular separation decreases (Large $\rightarrow$ Medium $\rightarrow$ Small), state-of-the-art methods degrade substantially, whereas NeCA++ remains consistent across datasets.

On ImageCAS, NeCA++ achieves the best Dice and IoU in the Large, Medium, and Small settings, with clear margins under Medium and Small angles (Dice 0.842 and 0.777, respectively); NeCA is unstable outside Orthogonal and NeRT-CA underperforms, while DeepCA preserves clDice in some cases but yields lower volumetric accuracy.
On ASOCA, our approach again outperforms all baselines, achieving Dice scores of 0.986, 0.941, and 0.932 from Large to Small angular separations; DeepCA attains competitive clDice but shows lower overlap accuracy and reduced stability as angular separation narrows.
On Synthetic RCA, similar trends are observed: NeCA performs strongly only under Orthogonal conditions but drops markedly under realistic angular constraints, and DeepCA achieves high clDice yet remains inferior in Dice and IoU, whereas our NeCA++ consistently delivers the highest overlap scores across Large, Medium, and Small settings, indicating improved robustness to limited angular diversity.

Across datasets, our proposed NeCA++ maintains high clDice even under narrow-angle settings, indicating preserved vascular connectivity, while achieving balanced improvements in both overlap and structural consistency.
Overall, the proposed multi-scale framework generalises and maintains stable performance as angular separation decreases, with advantages most pronounced under the clinically realistic Medium and Small angular separation settings. 
Because all experiments are performed at $80^3$ resolution due to memory constraints, the reported metrics may partly benefit from the coarser representation. Hence, we evaluate the upsampled reconstructions against $256^3$ ground-truth volumes, achieving Dice scores of 0.71 on ImageCAS and 0.78 on ASOCA and outperforming the strongest baseline (NeCA) by 30\% and 27\%, respectively.

\subsection{Ablation Study}
\label{sec:Ablation}

We evaluate the contributions of the proposed multi-stage optimisation and regularisations using four variants: NeCA++ without multi-stage and two regularisations (NeCA), NeCA++ without $L_{\text{Ray}}$ (W/o  $L_{\text{Ray}}$), NeCA++ without $L_{\text{bin}}$ (W/o  $L_{\text{bin}}$), and NeCA++ without multi-stage (W/o MS).
As shown in Table~\ref{tab:Ablation}, $L_{\text{Ray}}$ is the primary factor driving performance gains under small angular separations. The multi-stage strategy yields limited accuracy improvement but significantly accelerates optimisation, reducing the average training time from 632\,s to 58\,s per case. Using $L_{\text{Bin}}$ alone degrades performance because it amplifies outliers under sparse supervision. However, when combined with $L_{\text{Ray}}$, it further improves results by encouraging sharper binarisation, which enhances the outlier removal effect of the ray constraint. More importantly, it accelerates network convergence.
\begin{table*}[t]
\centering
\caption{Ablation results averaged over all datasets for each angular setting.}
\label{tab:Ablation}
\fontsize{8}{9}\selectfont
\setlength{\tabcolsep}{3pt}
\renewcommand{\arraystretch}{0.95}
\begin{tabular}{lcccccccccc}
\toprule
\multirow{2}{*}{Method}
& \multicolumn{3}{c}{Large}
& \multicolumn{3}{c}{Medium}
& \multicolumn{3}{c}{Small}
& \multirow{2}{*}{Time$\downarrow$} \\

\cmidrule(lr){2-4}
\cmidrule(lr){5-7}
\cmidrule(lr){8-10}

& Dice$\uparrow$ & IoU$\uparrow$ & clDice$\uparrow$
& Dice$\uparrow$ & IoU$\uparrow$ & clDice$\uparrow$
& Dice$\uparrow$ & IoU$\uparrow$ & clDice$\uparrow$
&  \\

\midrule

NeCA  & 0.587 & 0.418 & 0.466  
      & 0.732 & 0.585 & 0.662  
      & 0.636 & 0.473 & 0.569  
      & 633s \\

W/o  $L_{\text{Ray}}$ 
      & 0.502 & 0.339 & 0.394  
      & 0.688 & 0.524 & 0.608  
      & 0.587 & 0.422 & 0.523  
      & \textbf{57s} \\

W/o  $L_{\text{Bin}}$   
      & 0.925 & 0.873 & 0.868  
      & 0.860 & 0.803 & 0.868  
      & 0.839 & 0.752 & 0.833  
      & 86s \\

W/o  MS 
      & 0.931 & 0.882 & 0.920 
      & 0.887 & 0.822 & 0.877  
      & 0.859 & 0.768 & 0.856  
      & 632s \\

NeCA++  
      & \textbf{0.944} & \textbf{0.900} & \textbf{0.924}  
      & \textbf{0.899} & \textbf{0.827} & \textbf{0.891}  
      & \textbf{0.863} & \textbf{0.773} & \textbf{0.857}  
      & 58s \\

\bottomrule
\end{tabular}
\end{table*}

\section{Conclusion}

This paper presents NeCA++, a multi-stage neural implicit framework for 3D coronary artery reconstruction from two angiographic projections with narrow angular separation. By progressively refining spatial support and representation capacity, and incorporating novel regularisation, the proposed approach improves stability under sparse and narrow-angle separation. Extensive experiments over three datasets demonstrate consistent superiority over existing methods, particularly under clinically realistic non-orthogonal configurations. The framework achieves efficient reconstruction within approximately one minute per case, highlighting its practical potential for real-world interventional settings. 

%
%
%
%
\begin{credits}
\subsubsection{\ackname}
The authors acknowledge the use of the services/facilities of the Institute of Biomedical Engineering (IBME), Department of Engineering Science, University of Oxford.
The work of AB was supported by the Royal Society University Research Fellowship (grant no. URF{\textbackslash}R1{\textbackslash}221314) and the British Heart Foundation (BHF) Oxford Centre of Research Excellence (RE/24/130024).
The work of ML was supported by the Royal Society Enhanced Research Expenses Grant, awarded to AB.

\subsubsection{\discintname}
The authors have no competing interests to declare that are relevant to the content of this article.
\end{credits}

\bibliographystyle{splncs04}
\bibliography{Paper-5185}

\end{document}